\documentclass{elsart41}
\usepackage{graphicx}
\usepackage{amssymb}
\usepackage{times}
\newcommand{\chunks}{r}
\begin{document}

\begin{frontmatter}



\title{Hole doped Hubbard ladders}


\author[Gw]{H. Fehske\corauthref{Name1}},
\ead{fehske@physik.uni-greifswald.de}
\author[Er]{G. Hager},
\author[Er]{G. Wellein},
\author[Ma]{E. Jeckelmann}

\address[Gw]{Institut f\"ur Physik, Ernst-Moritz-Arndt-Universit\"at Greifswald, D-17487 Greifswald, Germany}
\address[Er]{Regionales Rechenzentrum Erlangen, Martensstr. 1, D-91058 Erlangen, Germany}
\address[Ma]{Institut f\"ur Physik, Johannes-Gutenberg-Universit\"at Mainz, D-55099 Mainz, Germany}

\corauth[Name1]{Corresponding author. Tel: +49 3834-864760, Fax: -864791}

\begin{abstract}
The formation of stripes in six-leg Hubbard ladders
with cylindrical boundary conditions is investigated 
for two different hole dopings, where the amplitude of 
the hole density modulation is determined in the limits
of vanishing DMRG truncation errors and infinitely long ladders. The 
results give strong evidence that stripes exist in the ground state of
these systems for strong but not for weak Hubbard couplings. The
doping dependence of these findings is analysed.
\end{abstract}

\end{frontmatter}

% main text

There is an ongoing controversial discussion about whether the ground
state of interacting doped lattice models in two dimensions like the
$t$-$J$ and the Hubbard model shows a charge modulation when subjected
to particular, e.g., cylindrical boundary conditions.

Recently, attention has turned to the two-dimensional
Hubbard model on $R\times L$--site ladders with  
local electron-electron repulsion $U$
and electron hopping $t$,
\begin{eqnarray}
H 
&=& -t \sum_{x,y,\sigma}
\left( c_{x,y,\sigma}^{\dag}c_{x,y+1,\sigma}^{\phantom{\dag}}
 + c_{x,y,\sigma}^{\dag}c_{x+1,y,\sigma}^{\phantom{\dag}}
 + \mbox{h.c.} \right )
\nonumber \\
&&+ U \sum_{x,y} n_{x,y,\uparrow}
n_{x,y,\downarrow}  \;,
\label{Hamiltonian}
\end{eqnarray}
where $x=1,\dots,R$ is the rung index and $y=1,\dots,L$ is the leg
index.  Cylindrical boundary conditions
(closed in the rung $y$ direction and open in the leg $x$
direction) are assumed. For $U=0$, this model describes a Fermi gas, 
obviously without stripes in the ground state.  Moreover, using 
renormalization group techniques, no
instability toward stripe formation has been found in the
weak-coupling limit $U \ll t$~\cite{Lin97}. 
In the strong-coupling limit $U \gg t$, 
the Hubbard model can be mapped onto a $t-J$ model with $J =
4t^2/U \ll t$, which exhibits stripes in the ground state, at least
on narrow ladders with $J \approx 0.35t$~\cite[and references therein]{WS03}.
An investigation of stripe formation in the Hubbard model
at different couplings and fillings could thus significantly 
improve our understanding of these structures.

In a recent density-matrix renormalization group (DMRG)
calculation White and Sca\-la\-pi\-no~\cite{WS03} have
shown that a narrow stripe appears in the ground state of a $7\times 
6$--site cluster for $U \geq 6t$\@.  For weaker couplings, the hole and
spin densities were interpreted as a broad
stripe. However, no finite-size scaling has been
performed, and the amplitude of the hole density modulation has not
been investigated systematically as a function of DMRG truncation
errors.

In this work we exclusively consider 6-leg ladders ($L=6$) with
$R=7\chunks$ rungs for $\chunks=1,\dots,4$.  Since we are interested
in the ground state of the hole-doped regime, we consider a system
with $N_1 = 4\chunks$ or $N_2 = 8\chunks$ holes doped in the
half-filled band, corresponding to $RL -N_1 = 38\chunks$ or
$RL-N_2=34\chunks$ electrons.  The average hole density is thus $n_1 =
N_1/RL = 4/42 \approx 0.095$ or $n_2 = N_2/RL = 8/42 \approx 0.190$,
respectively.
\begin{figure}
\centerline{\includegraphics*[width=0.9\columnwidth]{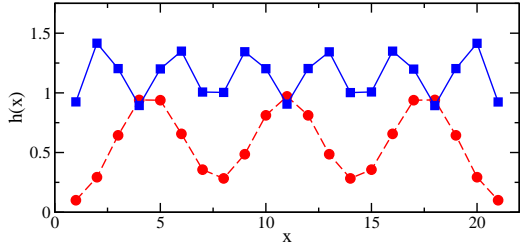}}
\caption{$y$-integrated hole density versus $x$ for $N_1=4r$ (circles)
and $N_2=8r$ (squares) on a $21\times 6$ ladder at $U=12t$.}\label{fig1}
\end{figure}

We employ a recently developed parallelised DMRG code~\cite{two},
keeping up to $m=8000$ density-matrix eigenstates per block 
for systems with up to $R\times
L= 126$ sites. We focus on the hole density $h(x,y) = 1-\langle
\hat{n}_{x,y,\uparrow}+\hat{n}_{x,y,\downarrow} \rangle$, 
where $\langle \dots \rangle$ represents the (DMRG)
ground-state expectation value.
A stripe is related to a hole density 
modulation in the leg direction,
\begin{equation}\label{hsum}
h(x) = \sum_{y=1}^{L} h(x,y) \; .
\end{equation}
Stripe ``signatures'' derived from the staggered spin density are
artifacts of the method and can be identified as such
\cite{one}\@.

The study of stripe structures requires a spectral analysis
of the hole density with respect to DMRG truncation errors and
finite-size effects. The spectral transform is defined as
\begin{equation}\label{hft}
H(k_x,k_y) = \sqrt{\frac{2}{L(R+1)}}
\sum_{x,y} \sin(k_x x) e^{i k_y y} h(x,y)
\end{equation}
with $k_x = z_x\pi/(R+1)$ for integers $z_x=1,\dots,R$ and $k_y = 2\pi
z_y/L$ for integers $-L/2 < z_y \leq L/2$.  In the converged DMRG
ground state we observe uniform behaviour of $h(x,y)$ along the rungs.
This implies that the spectral weight is concentrated at $k_y=0$\@.
Stripes appear as hole concentrations which are translationally
invariant along the rung ($y$) direction.  At a hole doping of
$N_1=4\chunks$, $r$ stripes show up in a $7 \chunks \times 6$ ladder.
When the doping is increased to $N_2=8\chunks$, the number of stripes
doubles (see Fig.~\ref{fig1}) and the structures become much less
pronounced.

The height and position of the maximum of the power spectrum [squared
norm of Eq.~(\ref{hft})], i.e., the dominant harmonic  
can be extrapolated to the limit of vanishing
DMRG truncation error (discarded weight $W_m$)\@. This is possible because 
for small  $W_m$, expectation values of operators are polynomials 
in $\sqrt{W_m}$\@. Indeed, we find a linear scaling of 
$H_{\mathrm{max}}=\max_{k_x}|H(k_x,0)|$ as $\sqrt{W_m}\to 0$ once
the transition to a striped state has occurred. 
For $U=12t$, the extrapolated values of $H_{\mathrm{max}}$ are 
finite. 
Thus we conclude that the hole density fluctuations
found on finite ladders
are not an artifact of DMRG truncation errors but a feature
of the true ground state for $U=12t$.
For smaller values of $U$, the fluctuation amplitude decreases until
$U\lesssim 4t$, and then stays close to zero if $U$ is further
reduced. 

\begin{figure}
\centerline{\includegraphics*[width=0.85\columnwidth]{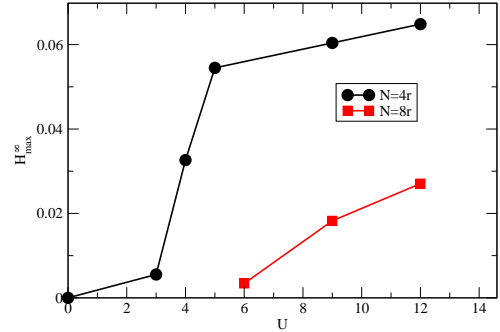}}
\caption{Amplitude of the dominant harmonic in the
	hole density modulation versus $U$, extrapolated to vanishing DMRG
	error and infinitely long ladders. Circles: $N=4\chunks$,
	squares: $N=8\chunks$\@.}\label{fig2}
\end{figure}

Eq.\ (\ref{hft}) implies that when the limit $R\to\infty$ is taken,
the amplitude of the dominant Fourier component in the hole density
modulation (\ref{hsum}) diverges linearly with
$\sqrt{R}$\@. Consequently, results for infinite ladder length can be
obtained by extrapolating $H_{\mathrm{max}}/\sqrt{R}$ as $R^{-1}\to
0$\@. Fig.~\ref{fig2} shows this limit for $N=N_1$ (circles)
and $N=N_2$ (squares)\@.

The following conclusions can be drawn from the data.  For small $U$,
stripe signatures observed in numerically determined ground states are
artifacts of the method and vanish when proper extrapolation
procedures are employed. Going from small to large $U$, there is a
crossover from a homogeneous to a striped state. With $N_1=4\chunks$, the
transition occurs with a rather steep slope at $U\approx
4t$\@. Increasing the doping to $N_2=8\chunks$ shifts the transition to
larger $U$ and makes it much smoother. Moreover, the number of stripes
in the ground state is doubled. The existence of a similiar transition 
in real two-dimensional strongly correlated electron systems would be of 
vital importance for the physics of layered high-$T_c$ cuprates.  

This work was partially funded by the Competence Network for Scientific
High Performance Computing in Bavaria (KONWIHR)\@. G.H.\ is indebted
to the HLRN Berlin staff.

\end{document}